\documentclass{iau}
\usepackage{graphicx}

\title["Stellar Prominences" on OB stars] 
{"Stellar Prominences" on OB stars\\to explain  wind-line variability}

\author[H. F. Henrichs \& N.P. Sudnik]   
{H.F. Henrichs$^1$ \and N.P. Sudnik$^2$}

\affiliation{$^1$Astronomical Institute Anton Pannekoek, University of Amsterdam, \\ Science Park 904,
1098 XH Amsterdam, Netherlands \\ email: {\tt h.f.henrichs@uva.nl} \\[\affilskip]
$^2$Sobolev Astronomical Institute, Saint Petersburg State University, \\ Universitetskij pr. 28, Staryj Peterhof, 198504, Saint Petersburg, Russia
\\email: {\tt snata.astro@gmail.com}}

\pubyear{2013}
\volume{302}  
\pagerange{xxx--xxx}
\jname{Magnetic fields throughout stellar evolution}
\editors{P. Petit, eds.}
\begin{document}

\maketitle

\begin{abstract}
Many O and B stars show unexplained cyclical
variability in their winds, i.e.\ modulation of absorption features on the rotational timescale, but not
strictly periodic over longer timescales. For these stars no dipolar magnetic
fields have been detected, with upper limits below 300 G. Similar cyclical
variability is also found in many optical lines, which are formed at the base
of the wind. We propose that these cyclical variations are caused by the presence of
multiple, transient, short-lived, corotating magnetic loops, which we call "stellar prominences".
We present a simplified model representing these prominences to explain the cyclical
optical wind-line variability in the O supergiant $\lambda$ Cephei.
Other supporting evidence for such prominences comes from the recent discovery of photometric variability in a
comparable O star, which was explained by the presence of multiple transient bright spots, presumably of
magnetic origin as well.

\end{abstract}

\firstsection 
\section{Introduction}

Since more than 30 years, spectroscopic UV observations from space have
shown that wind variability in massive O and B stars is a wide spread
phenomenon. This variability is not strictly periodic, but cyclic (like
sunspots) with often a dominant quasi period which scales with the estimated rotation period (days to weeks), or an integer fraction thereof
(e.g.\ \cite[Prinja and Howarth (1986)]{prinja86}, \cite[Kaper et al.\ (1996, 1997, 1999)]{kaper96, kaper97, kaper99}, \cite[Massa et al.\ (1995)]{massa95}, \cite[Prinja (1988)]{prinja88}, \cite[Henrichs et al.\ (1988)]{Henrichs88}). The underlying cause or trigger of this variability
is, however, unknown. Coordinated ground- and space-based studies show
that the major time-variable
wind features that are observed in the UV (the so-called DACs (discrete absorption components), must start from very near, or at the stellar surface (e.g.\ \cite[Henrichs et al., 1994]{henrichs94}, \cite[de Jong et al.,\ 2001]{dejong01}). The
presence of non-radial pulsations or bright magnetic star spots have
been suggested as a possible explanation (\cite[Henrichs et al., 1994]{henrichs94}, \cite[Cranmer and Owocki, 1996]{cranmer96}).

Pulsations have been found sofar only for a handful of O stars, mostly
by analyzing photospheric spectral line behavior (see \cite[Henrichs, 1999]{Henrichs99}),
but also from space-based photometry (\cite[Walker et al., 2005]{walker05}). Magnetic dipole
fields in such stars (except for the well-known class of chemically
peculiar stars) have been found since 1998, with the first magnetic O
star detection in 1999 ($\theta^1$ Ori C, \cite[Donati et al., 2002]{donati02}). Both phenomena are expected, however, to cause
strictly periodic variations, and are therefore unlikely the cause of the
observed cyclical behavior.

New very promising developments in this field are twofold.
First, theoretical studies show that in the sub-surface convective
layers in massive stars (\cite[Cantiello et al.,\ 2009]{cantiello09}), magnetic fields can be generated with a short
estimated turnover time (\cite[Cantiello and Braithwaite, 2011]{cantiello11}).  Second, high-precision space-based photometry
of the O giant $\xi$~Per showed rapid variations at the 1 mmag level, incompatible with the observed pulsations,
but compatible with the presence of a multitude of
corotating bright spots, which live only a few days (\cite[Ramiaramanantsoa et al.\ 2013]{ramansoara13}). These spots are
suggested to be of magnetic origin as described above, and which are
capable of triggering the wind variability in the form of DACs.

In this context, to understand the role of magnetic fields in O and
early B stars is a major challenge in massive star research. Here we focus on a simplified model to explain optical wind-line variability in the O supergiant $\lambda$ Cep, the behavior of which is representative for many other O stars. We conclude with an summarizing overall picture.

\section{Cyclical variability in {\boldmath $\lambda$} Cep O6I(n)fp}
\begin{figure}[b!]
\begin{center}
\includegraphics[width=\columnwidth]{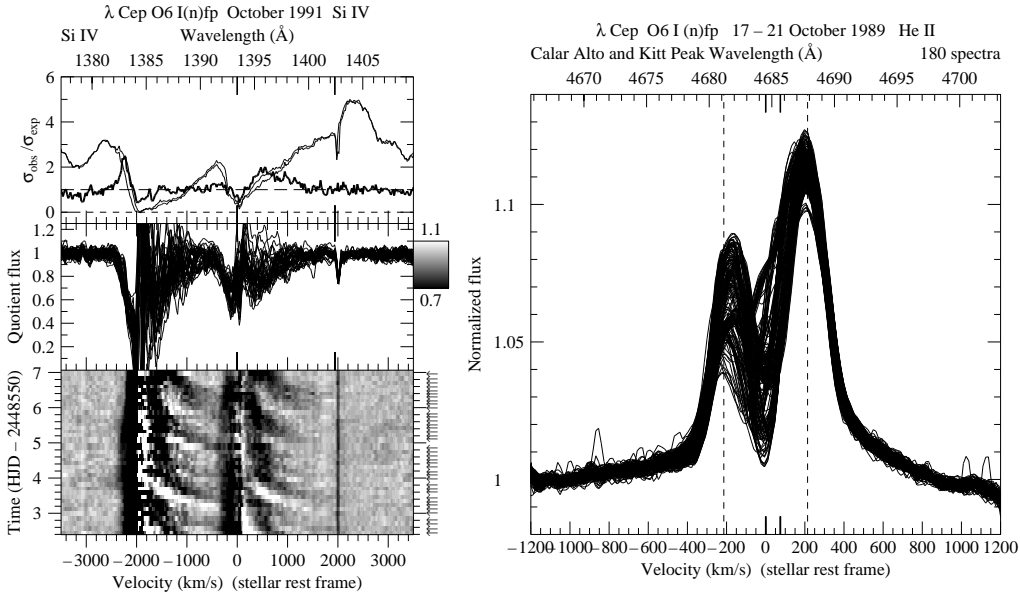}
\caption{{\it Left:} Cyclical DACs in the Si IV doublet quotient UV spectra of $\lambda$ Cep as observed in 1991 (from \cite[Kaper et al.\ (1999)]{kaper99}, Fig. 27). {\it Right:} Cyclical He II 4686 variability over 4 days as observed in 1989 (Kitt Peak and Calar Alto). Significant changes occur in 15 min.}
\label{fig1}
\end{center}
\end{figure}

\begin{figure}[h!]
\begin{center}
\includegraphics[width=\columnwidth]{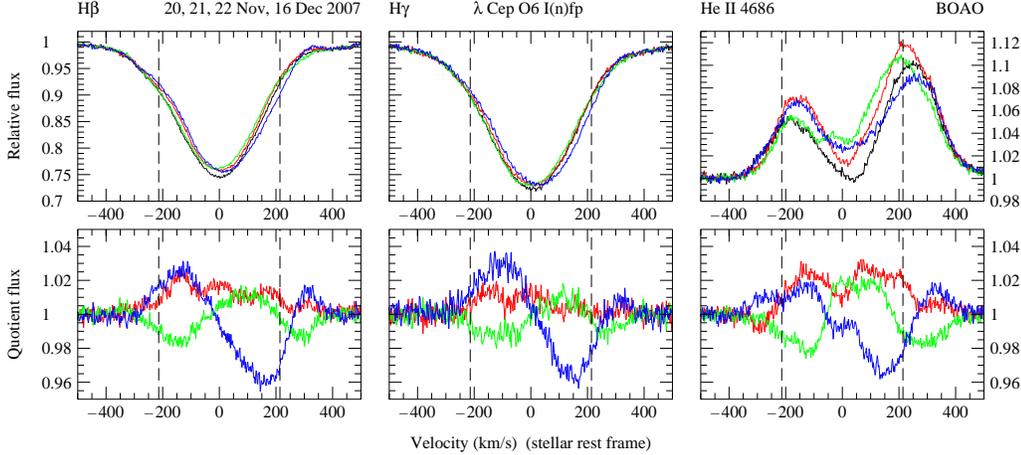}
\caption{{\it Top row:} Nightly averages of H$\beta$, H$\gamma$ and He II 4686 lines of the 2007 BOAO dataset.  {\it Bottom row:} Quotient spectra of the same lines of subsequent nights. Note that although the variability  of the nightly averages of the H$\beta$ and H$\gamma$ lines are much more subtle than of the He II line, the quotient spectra are qualitatively rather similar. Also note that the variability clearly extends beyond the $\mid $$v$sin$i$$\mid$ range.}
\label{fig2}
\end{center}
\end{figure}

The bright runaway star O6I(n)fp star $\lambda$ Cep ($v$sin$i \simeq 214$ km/s, log($L/L_{\odot})\simeq 5.9$, $T_{\rm eff} \simeq 36000$ K, $R \simeq 16 R_{\odot}$, $M \simeq 60 M_{\odot}$, \cite[Markova et al.\ 2004]{markova04}) is a nonradial pulsator ($l = 3, P= 12.3$ h; $l = 5, P = 6.6$ h, \cite[de Jong et al.\ 1999]{dejong99}), and shows cyclical DACs in the UV resonance lines. Rapid variability have been observed in the He II emission line in 1989 (Fig.\,\ref{fig1}), as confirmed in later studies at BOAO (Korea) in 2007, used for the analysis below, and in 2012. The dominant period in the UV and optical lines is $\simeq 2$ d. Only redward moving NRP features have been observed, implying an inclination angle greater than, say, $50^{\circ}$. With the adopted radius, the likely rotation period is then $\simeq 4$ d.
We also found that the H, He I and other He II lines behave remarkably similarly. This becomes only apparent by considering quotient spectra of subsequent nights (see Fig.\ \ref{fig2}). We note that we found this covariability in many spectral lines also for other O stars.
This variability extends clearly beyond the $v$sin$i$ limits, which implies that the corotating emitting gas sticks out above the stellar surface. This lead us to suggest the term "stellar prominences" to characterize this phenomenon.

\vspace*{-0.6 cm}

\begin{figure}[b!]
\begin{center}
\includegraphics[width=\columnwidth]{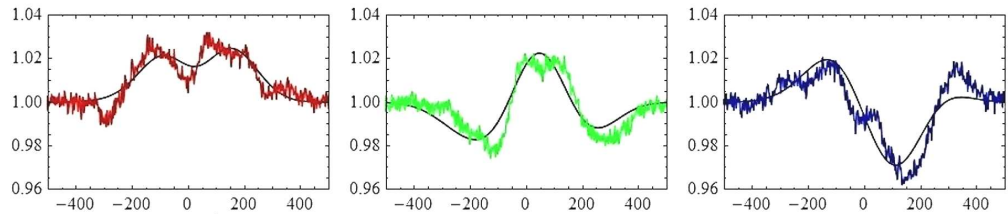}
\caption{Illustrative model fits to the three He II quotient spectra of the lower right panel of Fig.\ \ref{fig2}. Six blobs were needed at a $4.07$ day rotation period and $i = 55^{\circ}$. The fit may not be unique, but at least constrains the minimum number of spots and the stellar rotation parameters. }
\label{fig3}
\end{center}
\end{figure}

\section{A simplified model for stellar prominences}
As the observed optical profile changes requires emitting gas above the surface, we consider the most simplified model to represent a prominence by a sphere (radius $r \simeq 0.2 R_{*}$, optical depth $\tau$) corotating and touching the surface. The corresponding line profile in velocity space is then given by 
$F(v)=\exp\left[-A\tau e^{-\frac{1}{2}\left(\frac{v-v_0}{w}\right)^2}\right]\exp\left[(\pi r^2 - A)\tau e^{-\frac{1}{2}\left(\frac{v-v_0}{w}\right)^2}\right]$
in which $0<A(t)\leq \pi r^2$ takes into account the transient and eclipsing geometry, in analogy with exoplanet analysis (including partial eclipses). Each model blob corotates with projected velocity $v_0$ and has profile width $w$.  The procedure is to adopt a fixed inclination angle (matching $v$sin$i$ and $R_*$) and put a number of blobs around the star to fit the first quotient, which is determined by an assumed rotation period. A least-squares fit gives the best parameters. The remaining quotient spectra should fit correctly, only if the rotation period is correct. This will constrain the rotation period for the adopted stellar parameters. The best fit to the quotients of the He II line (lower right panel of Fig.\ \ref{fig2}) is shown in Fig.\ \ref{fig3}. Fitting quotient profiles of other spectral lines, also in other datasets, is work in progress.

\section{Summarizing picture}
The overall visualized picture is as follows. The conjectured prominences are located above the corotating magnetic bright spots. Close to the star the enhanced mass flow above a magnetic spot is for simplicity envisaged to be similar to above active regions around sunspots, here labeled as stellar prominences. The mass outflow is stronger above these spots, locating the footpoints of DACs which arise because of the velocity plateau in this outflow, giving enhanced absorption. This part of the wind moves slower then the surrounding wind as it is overloaded because it receives relatively less driving force.  The bright spots give rise to small ($\simeq 1$ mmag) photometric variability, as observed in $\xi$~Per, and predicted for $\lambda$ Cep and other OB stars with DACs. The magnetic spots are bright because of the higher temperature in the deeper layers where the energy is transported by radiation (as opposed to sunspots where the energy transport is by convection). These magnetic fields are the result of the subsurface convection as described by \cite[Cantiello and Braithwaite (2011)]{cantiello11}. The lifetime of these generated fields is determined by the relatively short turnover time of this subsurface layer. The maximum strength of the field is estimated by equipartition considerations. The strength of these fields imply the magnetic confinement parameter to be around unity (\cite[ud-Doula and Owocki, 2002]{uddoula02}). The detection, with current instrumentation, of such magnetic field configurations of many magnetic spots distributed over the surface has been studied by \cite[Kochukhov and Sudnik (2013)]{kochukhov13}. Partial cancellation effects will occur. Such studies, together with observed upper limits, can constrain the number of spots and their distribution.\\
Future space studies of photometric variability, coordinated with UV and ground-based spectroscopy are needed to test the picture sketched above.
\\

{\sl Acknowledgements.} We thank Stan Owocki and Marianne Faurobert for constructive and insightful discussions.


\begin{thebibliography}{}

\bibitem[Cantiello et al.\ (2009)]{cantiello09}
{Cantiello, M., Langer, N., Brott, I., de Koter, A., Shore, S.N., et al.}, 2009
\textit{A\&A}, 499, 279

\bibitem[Cantiello and Braithwaite (2011)]{cantiello11}
{Cantiello, M., \& Braithwaite, J.}, 2011,
\textit{A\&A}, 543, 140

\bibitem[Cranmer and Owocki, 1996]{cranmer96}
{Cranmer, S.R. \& Owocki, S.A.}, 1994,
\textit{ApJ}, 462, 469

\bibitem[de Jong et al.\ (1999)]{dejong99}
{de Jong, J.A., Henrichs, H.F., Schrijvers, C., Gies, D.R., et al.}, 1999,
\textit{A\&A}, 345, 172

\bibitem[de Jong et al.\ (2001)]{dejong01}
{de Jong, J.A., Henrichs, H.F., Kaper, L., Nichols, J.S., et al.}, 2001,
\textit{A\&A}, 368, 601

\bibitem[Donati et al.\ (2002)]{donati02}
{Donati, J.-F., Babel, J., Harries, T. J., et al.}, 2002,
\textit{MNRAS}, 333, 55

\bibitem[Henrichs (1999)]{henrichs99}
{Henrichs, H.F.}, 1999,
\textit{Lecture Notes in Physics, Berlin Springer Verlag}, 523, 305

\bibitem[Henrichs et al.\ (1988)]{henrichs88}
{Henrichs, H. F., Kaper, L. \& Zwarthoed, G. A. A.}, 1988
\textit{In ESA, Proc.\ a Decade of UV Astronomy with the IUE Satellite}, Volume 2, 145

\bibitem[Henrichs et al.\ (1994)]{henrichs94}
{Henrichs, H.F., Kaper, L., \& Nichols, J.S.}, 1994,
\textit{A\&A}, 285, 565

\bibitem[Kaper et al.\ (1996)]{kaper96}
{Kaper, L., Henrichs, H.F., Nichols, J.S., et al.}, 1996,
\textit{A\&ASS}, 116, 257

\bibitem[Kaper et al.\ (1996)]{kaper96}
{Kaper, L., Henrichs, H.F., Fullerton, A.W., et al.}, 1997,
\textit{A\&A}, 327, 281

\bibitem[Kaper et al.\ (1999)]{kaper99}
{Kaper, L., Henrichs, H.F., Nichols, J.S., \& Telting, J.H.}, 1999,
\textit{A\&A}, 344, 231

\bibitem[Kochukhov and Sudnik (2013)]{kochukhov13}
{Kochukhov, O. \& Sudnik, N.}, 2013,
\textit{A\&A}, 554, 93

\bibitem[Markova et al.\ (2004)]{markova04}
{Markova, N., Puls, J., Repolust, T. \& Markov, H.}, 2004,
\textit{A\&A}, 413, 693

\bibitem[Massa et al.\ (1995)]{massa95}
{Massa, D., Fullerton, A.W., Nichols, J.S., et al.}, 1995,
\textit{ApJ}, 452, L53

\bibitem[Prinja (1988)]{prinja88}
{Prinja R.K.}, 1988,
\textit{MNRAS} 231, 21P

\bibitem[Prinja and Howarth (1986)]{prinja86}
{Prinja R.K.  \& Howarth I.D.}, 1986,
\textit{ApJS} 61, 357

\bibitem[Ramiaramanantsoa et al., (2013)]{Ramiaramanantsoa13}
{Ramiaramanantsoa, T., Moffat, A.F.J., Chen\'{e}, A.-N., Desforges, S.,
Richardson, N. D., Henrichs, H.F., Guenther, D.B., Kuschnig, R., et al.}, 2013,
\textit{MNRAS} submitted

\bibitem[ud-Doula and Owocki (2002)]{uddoula02}
{ud-Doula, A. \& Owocki, S.P.}, 2002,
\textit{ApJ}, 576, 413

\bibitem[Walker et al.\ (2005)]{walker05}
{Walker, G. A. H., Kuschnig, R., Matthews, J. M., et al.}, 2005,
\textit{ApJL}, 623, L145

\end{thebibliography}
\end{document}